\documentclass[a4paper]{article}

\usepackage{INTERSPEECH2020}

\usepackage[caption=false]{subfig}
\usepackage{tikz}
\usepackage[ruled,vlined,linesnumbered]{algorithm2e}

\title{The UPC Speaker Verification System Submitted to VoxCeleb Speaker Recognition Challenge 2020 (VoxSRC-20)}
\name{Umair Khan and Javier Hernando}
\address{TALP Research Center, Department of Signal Theory and Communications, \\
Universitat Politecnica de Catalunya Barcelona, Spain}
\email{\{umair.khan, javier.hernando\}@upc.edu}

\begin{document}

\maketitle
\begin{abstract} 
This report describes the submission from Technical University of Catalonia (UPC) to the VoxCeleb Speaker Recognition Challenge (VoxSRC-20) at Interspeech 2020. The final submission is a combination of three systems. System-1 is an autoencoder based approach which tries to reconstruct similar i-vectors, whereas System-2 and -3 are Convolutional Neural Network (CNN) based siamese architectures. The siamese networks have two and three branches, respectively, where each branch is a CNN encoder. The double-branch siamese performs binary classification using cross entropy loss during training. Whereas, our triple-branch siamese is trained to learn speaker embeddings using triplet loss. We provide results of our systems on VoxCeleb-1 test, VoxSRC-20 validation and test sets. 
\end{abstract}

\section{Introduction}
This report explains the submission from Technical University of Catalonia (UPC) to the VoxCeleb Speaker Recognition Challenge (VoxSRC-20), which is held in conjunction with Interspeech 2020. The VoxSRC-20 is the second edition of the speaker recognition challenge held by VoxCeleb team. The challenge has four separate tracks, three of them are dedicated to speaker verification task and one to speaker diarization. The UPC system was submitted to Track 3 which is a closed self-supervised speaker verification task. 

The submitted result is a combination of three systems. In system-1 we trained an autoencoder to reconstruct neighbor i-vectors, rather than the same training i-vectors. After the training, we extract speaker vectors for the testing i-vectors, which are referred to as autoencoder vectors or shortly ae-vectors \cite{interspeech-2019}. In system-2 and -3 we trained two siamese networks \cite{siamese-1-1994}, i.e., double-branch and triple-branch, which consist of two and three branches, respectively \cite{interspeech-2020}. Each branch is composed of a Convolutional Neural Network (CNN) encoder, inspired by the VGG architecture \cite{vgg-2014, voxceleb-1, voxceleb-2}. The double-branch network is trained as a binary classifier by minimizing binary cross entropy loss. After training, we obtain decision scores for the speaker verification trials from the output of the network. The triple-branch network is trained by minimizing triplet loss \cite{triplet-image-wang-2014, triplet-image-schroff-2015}. In the testing phase, we extract speaker embeddings for the test data using any branch of the triple-branch network, which are scored using cosine scoring. 

The rest of the report is organized as follows. Section \ref{Proposed Method} explains the three systems in detail. Section \ref{Experimental setup and database} describes the experimental setup and results. Finally, conclusions are drawn in Section \ref{Conclusions}. 

\section{Proposed method}
\label{Proposed Method}

\begin{figure}[t]
\centerline{\includegraphics[height=0.195\textheight,width=0.47\textwidth]{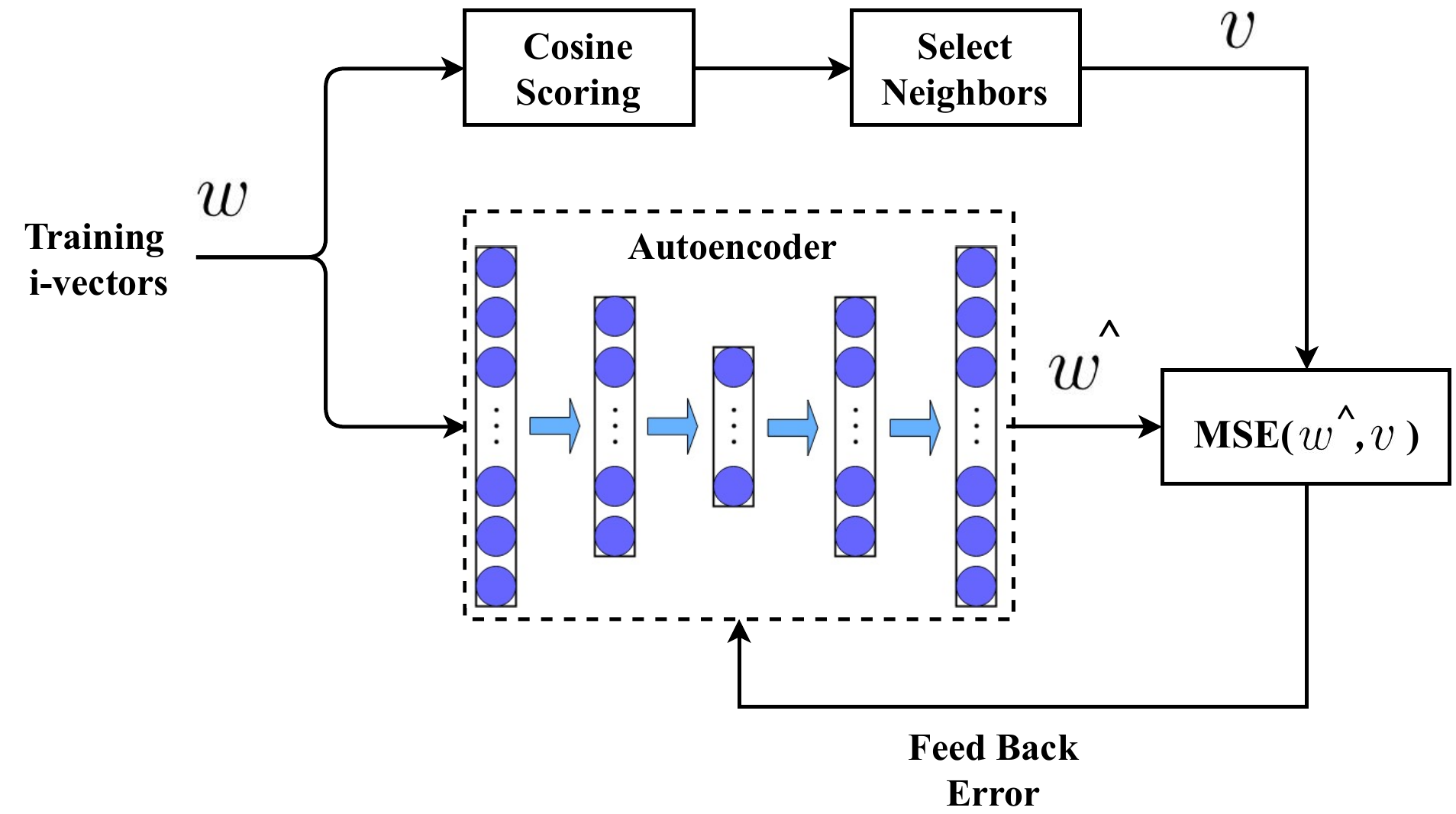}}
\caption{Architecture of System-1: Autoencoder}
\label{fig:ae}
\end{figure}

\subsection{System-1: Nearest Neighbor Autoencoder}
\label{Autoencoder Training}

System-1 is an autoencoder which is trained by minimizing the loss function : $MSE(w\string^, v)$, as shown in Figure \ref{fig:ae}, where $v$ is a similar i-vector to $w$ and $w\string^ = decoder(encoder(w))$. For every i-vector we select a set of similar i-vectors as neighbor i-vectors. All the training i-vectors are scored among each other using cosine scoring. We select the top $k$ i-vectors with highest scores as neighbor i-vectors. In the testing phase, we transform the test i-vectors into new speaker vectors by extracting at the output of the autoencoder. These are referred to as autoencoder vectors or shortly ae-vectors. Details can be found in \cite{interspeech-2019, icassp-2020-from-ieee}. 

\begin{figure}[t]
\centerline{\includegraphics[height=0.35\textheight,width=0.33\textwidth]{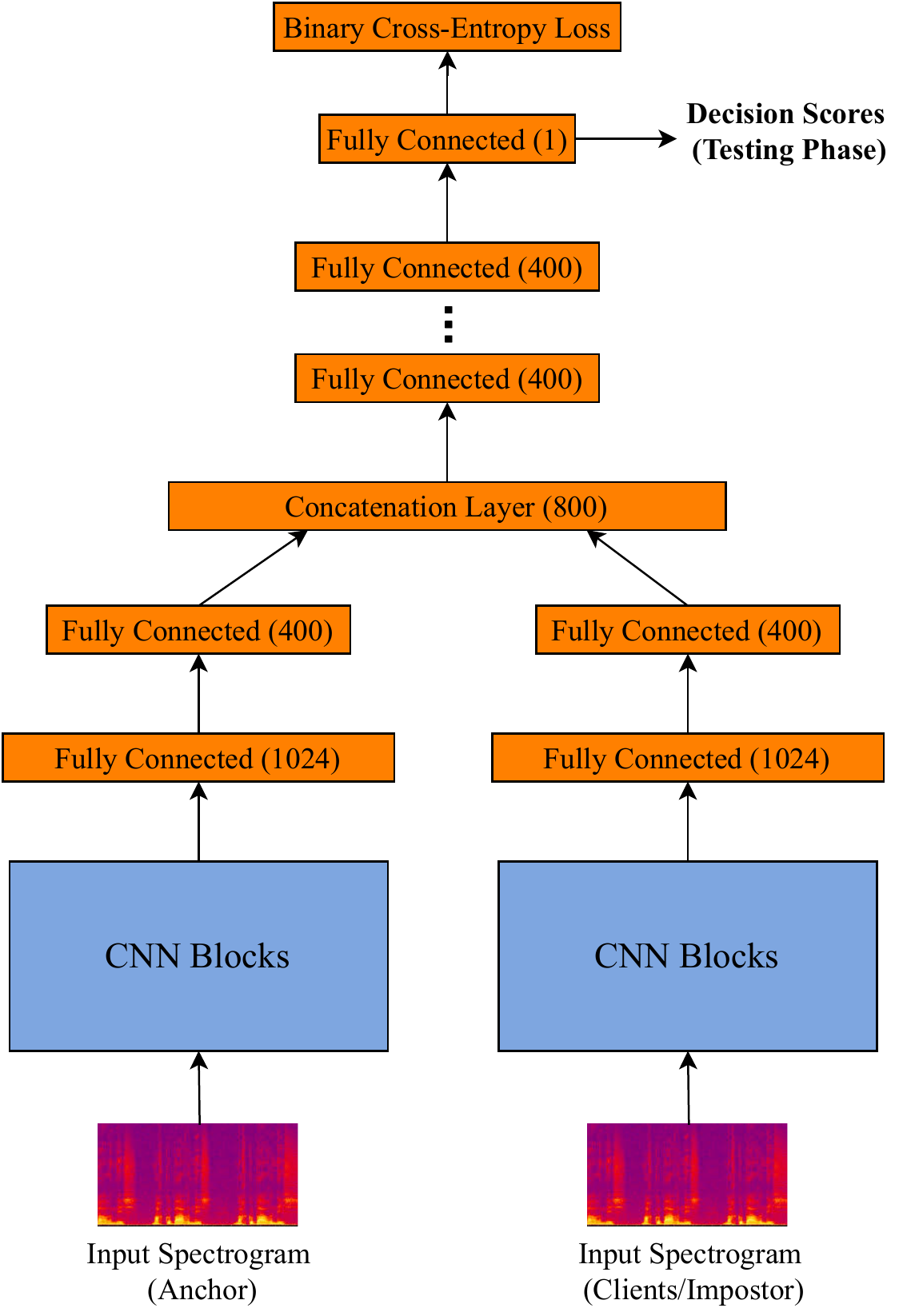}}
\caption{Architecture of System-2: Double-branch siamese.}
\label{fig:siamese}
\end{figure}

\subsection{System-2: Double-branch siamese network}
\label{Double-branch siamese network}

\begin{figure}[t]
\centering
\includegraphics[height=0.27\textheight,width=0.46\textwidth]{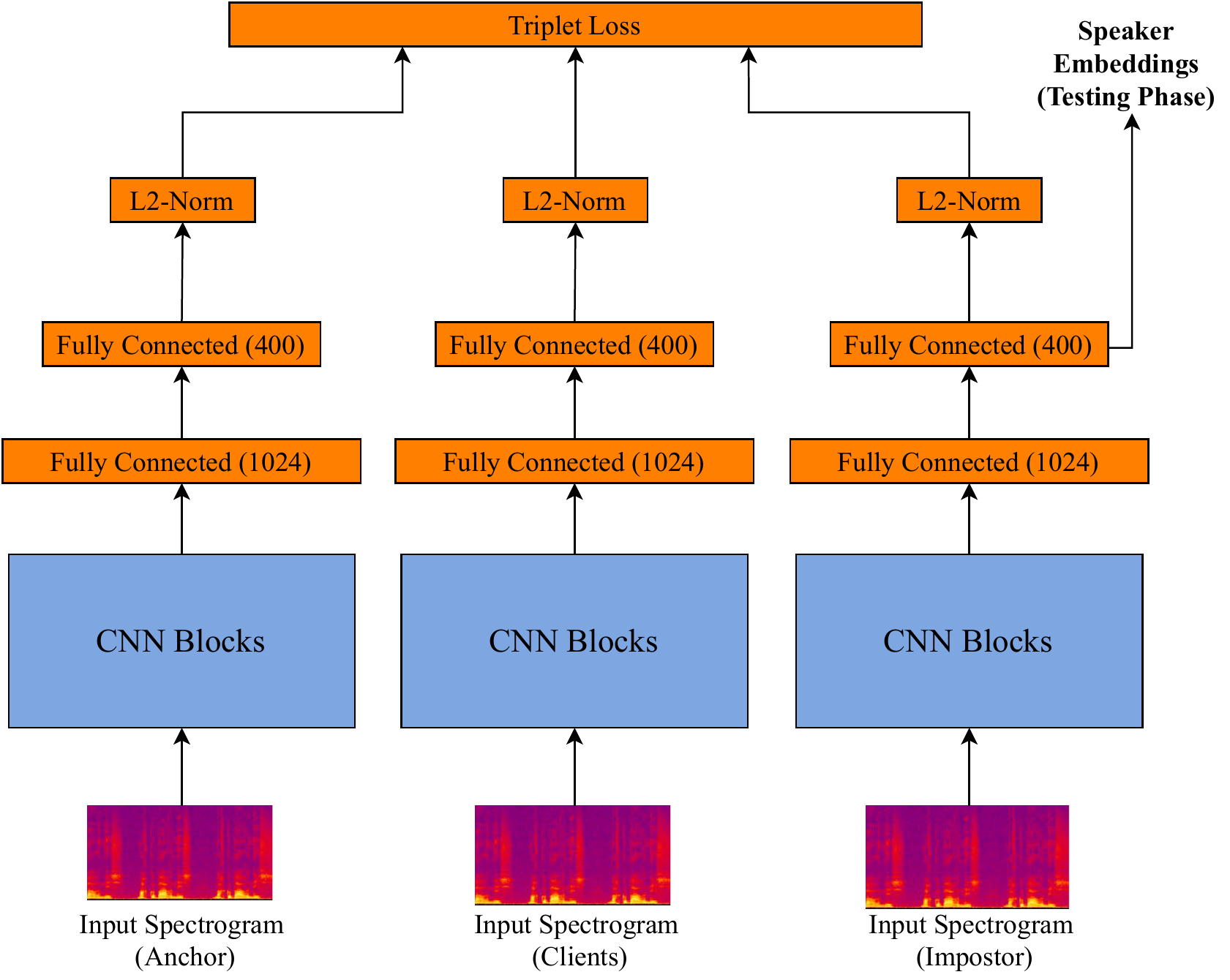}
\caption{Architecture of System-3: Triple-branch siamese.}
\label{fig:triplet}
\end{figure}

Figure \ref{fig:siamese} shows the architecture of our double-branch siamese network. There are two identical branches, i.e., the CNN encoder. Mel-spectrogrm features of a training pair of an anchor along with a client or an impostor sample is fed into the network. The two branches share weights and biases with each other. The outputs of the two branches are concatenated which is followed by five Fully Connected (FC) layers. The last layer is connected to the binary class labels, i.e., 1/0, indicating if the anchor sample is paired by a client/impostor sample, respectively. During training, binary cross-entropy loss is minimized. A pair of reference and test utterances, which is involved in an experimental trial, is fed into the network. The decision scores are obtained directly from the output \cite{interspeech-2020}. 

\subsection{System-2: Triple-branch siamese network}
\label{Triple-branch siamese network}

A block diagram of our triple-branch siamese network is shown in Figure \ref{fig:triplet}. As indicated by the name, there are three branches, each of which is the CNN encoder followed by l2-normalization layer. Each branch is fed by the Mel-spectrogrm features of a training pair of an anchor along with a client and an impostor sample. The CNN encoder encodes the Mel-spectrogram inputs into a vector based representation i.e., speaker embeddings. After the l2-normalization layer, triplet loss \cite{triplet-image-wang-2014} is computed between the embeddings of anchor, client and impostor samples. Once the network is trained, we extract speaker embeddings using any CNN encoder branch of the network \cite{interspeech-2020}. 

\subsection{Selection of clients and impostor samples}
\label{Selection of clients and impostor samples}

Our siamese networks are trained using pairs of training samples, i.e., anchor, client and impostor. Since we do not use speaker labels, we generate the training pairs in a self-supervised manner \cite{interspeech-2020}. The selection process of client and impostor samples is carried out in the i-vector space using two non-overlapping subsets of VoxCeleb-2 database, i.e., $A$ and $B$. We assume that the speakers in one subset do not appear in the other. We extract i-vectors for all the utterances in both the subsets. Then, all the i-vectors in $A$ are scored among each other using cosine scoring. For every i-vector in $A$ we select $k$ number of similar i-vectors as potential client i-vectors. The potential clients are subjected to a $threshold$ and final client i-vectors are obtained. 

After this, we score all the i-vectors in $A$ with those in $B$, using cosine scoring. For every i-vectors in $A$, we select $k$ number of i-vectors from $B$ according to the cosine scores. Since, we assume that the speakers in $A$ does not appear in $B$, the selected i-vectors are assumed as the potential impostors, which are subjected to a threshold. In this way, every i-vector in $A$ has been assigned $k$ client and $k$ impostor i-vectors. For the double-branch network we make training pairs of two, i.e., [anchor, client] and [anchor, impostor], for which the binary labels are 1 and 0 respectively. Whereas for the triple-branch network we make tuples of three samples, i.e., [anchor, client, impostor]. This process is carried out in i-vector space while the actual inputs to the networks are Mel-spectrogram features \cite{interspeech-2020}. 

\subsection{CNN encoder}
\label{Cnn encoder}

The CNN encoder is inspired by the VGG architecture \cite{vgg-2014}, recently adapted for speaker verification in \cite{voxceleb-1, voxceleb-2, interspeech-2020}. It consists of three blocks, each containing two convolutional and one maxpooling layer as shown in Table \ref{tab:cnn}. The blocks are followed by self attention pooling (SAP) layer \cite{self-attention-2018}, and two FC layers. 

\begin{table}[t]
\caption{Architecture of the VGG based CNN Encoder.}
\begin{tabular}{|c|c|c|c|c|c|}
\hline
Layer     & Size & In dim & Out dim & Stride & Feat size \\ \hline
conv1-1   & 3x3  & 1      & 128     & 1x1    & 80xN      \\ \hline
conv1-2   & 3x3  & 128    & 128     & 1x1    & 80xN      \\ \hline
mpool-1   & 2x2  & -      & -       & 2x2    & 40xN/2    \\ \hline
conv2-1   & 3x3  & 128    & 256     & 1x1    & 40xN/2    \\ \hline
conv2-2   & 3x3  & 256    & 256     & 1x1    & 40xN/2    \\ \hline
mpool-2   & 2x2  & -      & -       & 2x2    & 20xN/4    \\ \hline
conv3-1   & 3x3  & 256    & 512     & 1x1    & 20xN/4    \\ \hline
conv3-2   & 3x3  & 512    & 512     & 1x1    & 20xN/4    \\ \hline
mpool-3   & 2x2  & -      & -       & 2x2    & 10xN/8    \\ \hline
SAP & -    & N/8    & 1       & -      & 512x10    \\ \hline
fc-1      & -    & 1      & 1       & -      & 1024    \\ \hline
fc-2      & -    & 1      & 1       & -      & 400    \\ \hline
\end{tabular}
\label{tab:cnn}
\end{table}

\section{Experiments and results}
\label{Experimental setup and database}

The training was performed on the development partition of VoxCeleb-2 database \cite{voxceleb-2}. The evaluation was performed on the test partition of VoxCeleb-1 \cite{voxceleb-1}, and the VoxSRC-20 validation and test sets. The autoencoder of System-1, is a fully connected feed forward network which consists of 3 hidden layers. The encoder and decoder parts are symmetrical. The hidden layer 1 and 3 have 300 neurons each, while hidden layer 2 consists of 200 neurons. The input and output layers consist of 400 neurons each. The autoencoder training was carried out for 100 epochs using Stochastic Gradient Descent (SGD) optimizer. All the layers of the autoencoder used ReLU activation except the last layer which used linear activation. The learning rate was set to 0.01 with a decay of 0.0002 and the batch size was set to 100. 

For the clients and impostor selection in Systems-2 and System-3, (since we participated in a closed track of the challenge) we split VoxCeleb-2 into two parts to generate subsets $A$ and $B$ as discussed in Section \ref{Selection of clients and impostor samples}. (For an open track, one could use two separate datasets, for instance, VoxCeleb-1 as $A$ and VoxCeleb-2 as $B$). Mel-spectrograms of 80 dimensions were computed, of which a randomly selected window of length $N = 350$ was input to the networks. All the CNN and fully connected layers were activated using ReLU function, whereas the last layer of the double-branch network used sigmoid activation. The training was carried out using Adam optimizer. The fusion scores were obtained using the following expression: 
\begin{equation}
Scores = ((S_1 \times \alpha)  + (S_2 \times (1-\alpha)) \times \beta) + (S_3 \times (1-\beta))
\end{equation}
\noindent Where $S_1$, $S_2$ and $S_3$ are the individual scores obtained using System-1, 2 and 3, respectively. The values of $\alpha$ and $\beta$ were tuned on the VoxSRC-20 validation set, and were set to 0.30 and 0.79, respectively. 

\begin{table}[t]
\renewcommand{\arraystretch}{1.12}
\caption{EER in \% and DCF obtained for best values of $k$, on VoxCeleb-1 test set.}
\begin{center}
\begin{tabular}{cccc}
\hline \hline
\textbf{Approach} & \textbf{k} & \textbf{EER(\%)} & \textbf{DCF}\\
\hline \hline
System-1: Autoencoder \cite{interspeech-2019}  &  15 &   10.20 & 0.8066 \\
\hline 
System-2: Double-branch \cite{interspeech-2020} &  10	&	6.90 & 0.7681 \\
\hline 
System-3: Triple-branch \cite{interspeech-2020} &  10	&	6.95 & 0.6606 \\
\hline \hline 
Fusion of the above  &  -  &	5.58 & 0.5452 \\
\hline \hline
\end{tabular}
\label{tab:eer}
\end{center}
\end{table}

\begin{table}[t]
\renewcommand{\arraystretch}{1.12}
\caption{EER in \% and DCF on VoxCeleb-1 test, VoxSRC-20 validation and test sets, using fusion of the three systems.}
\begin{center}
\begin{tabular}{ccc}
\hline \hline
\textbf{Approach} & \textbf{EER(\%)} & \textbf{DCF} \\
\hline \hline
VoxCeleb-1 (test set)  &  5.58  &	0.5452 \\
\hline
VoxSRC-20 (validation set) &  14.30  &	0.6921  \\
\hline
VoxSRC-20 (test set)  &  14.71  &	0.7514  \\
\hline \hline
\end{tabular}
\label{tab:eer-2}
\end{center}
\end{table}

Table \ref{tab:eer} shows the EER in \% and DCF obtained for the best values of $k$ on VoxCeleb-1 test set (also reported in \cite{interspeech-2019, interspeech-2020}). Each row contains the results obtained using the individual systems. From the Table we can see that the double-branch siamese network has achieved the best EER of 6.90\% which is almost similar to that of the triple-branch network. Moreover, if we perform a score level fusion of all the three systems, we obtain an EER and DCF of 5.58\% and 0.5452, respectively. 

Table \ref{tab:eer-2} shows the EER and DCF obtained on VoxSRC-20 validation and test sets. The results were obtained using the fusion of the individual scores of the three systems. Our submission ranked 3rd in the VoxSRC-20 competition (Track 3). 

\section{Conclusions}
\label{Conclusions}
This report describes the submission of Technical University of Catalonia (UPC) to the VoxCeleb Speaker Recognition Challenge (VoxSRC-20) at Interspeech 2020. The submitted result is a combination of three systems, i.e., autoencoder, double- and triple-branch siamese networks based on Convolutional Neural Network (CNN). Brief analysis of the proposed systems (individual and in fusion) is provided on VoxCeleb-1 test set. The fusion results for VoxSRC-20 validation and test sets have also been reported. 

\section{Acknowledgements}

This work was supported by the project PID2019-107579RB-I00 / AEI / 10.13039/501100011033

\bibliographystyle{IEEEtran}
\bibliography{My_All_Bibliography}

\end{document}